\documentclass[11pt]{amsart}

\usepackage{graphicx}
\usepackage{dcolumn}
\usepackage{bm}
\usepackage{amssymb,amsmath}
\usepackage{doublespace}

\def\ep{\epsilon}

\def\W{\Omega}
\def\w{\omega}
\def\ep{\epsilon}
\def\p{\varrho}
\def\a{\alpha}
\def\be{\beta}
\def\*{\cdot}
\def\D{\Delta}
\def\d{\partial}
\def\de{\delta}
\def\he#1{#1^\dagger}
\def\ba#1{\overline{#1}}
\def\v#1{\mathbf{#1}}

\begin{document}


\title{Application of quantum inequalities to quantum optics}

\author{Piotr Marecki}


\begin{abstract}

We establish a connection between quantum inequalities, known from quantum field theory
on curved spacetimes, and the degree of squeezing in quantum-optical experiments. We
prove an inequality which binds the reduction of the electric-field fluctuations to their
duration. The bigger the level of fluctuation-suppression the shorter its duration. As
an example of an application of this inequality is the case of squeezed light whose phase
is controlled with $1\%$ accuracy for which we derive a limit of $-15dB$ on the allowed
degree of squeezing.

\end{abstract}

\maketitle

\section{Introduction}

In quantum field theory the normal-ordered energy density does not need to be positive.
In other words the expectation value of the energy density at a point $x$
\begin{displaymath}
  \p(x)=\langle :T_{00}(x):\rangle_S,
\end{displaymath}
for certain states $|S\rangle$  of the quantum field, can be arbitrarily negative. Let
us give a simple example, consider the following state
\begin{equation}\label{state}
  |S\rangle=N(|\Omega\rangle + \ep |fg\rangle),
\end{equation}
which is a superposition of the vacuum state $|\Omega\rangle$ and two particle state
$|fg\rangle$ \footnote{N is a normalization factor, the particles are described by
wavepackets $f(p)$ and $g(p)$ i.e. $|fg\rangle= \int d^3p \ d^3k \ \ba{f(p)} \ba{g(k)}
\he a(p) \he a(k) |\Omega\rangle$, where for simplicity the polarization was
disregarded.}. A calculation shows that the energy density, at a certain point $x$,
contains two, generally non-vanishing, terms
\begin{displaymath}
  \p(x)=\ep A(x)+\ep^2 B(x)
, \quad B(x)\geqslant 0
\end{displaymath}
Evidently we can choose the sign and the magnitude of $\ep$ in
such a way, that $\p(x)$ becomes negative at the point $x$.

Since the time that the appearance of negative energies in quantum field theory has been
recognized we have learned a great deal about this phenomenon. Above all, the integrated
energy density (i.e. the total energy)  must always be positive. Therefore the negative
energy densities are expected to be present only locally and are suppressed by positive
ones in adjacent regions. Moreover, there is an important question about whether negative
energy densities can be present at a certain point $x$ for a longer time. The fact that
they cannot was discovered by L.Ford \cite{ford}. Physically, the longer the time of
measurement  the less negative the energy becomes. A whole branch of theoretical physics
grew out of this pioneering work. The so-called quantum inequalities have been proven
with  great generality\cite{verch,fewster} for various types of fields, eg.
electromagnetic \cite{pfenning}, Dirac \cite{few_verch}, even in the situation where the
fields propagate in a curved spacetime (which is far more difficult than anything we
shall present here).

On the other hand, curiously, the type of states (\ref{state}) have  recently become a
standard tool in quantum optics. Known as squeezed states they arise in the process of
parametric down conversion\cite{kimble_sq_gen} where an incident photon is converted in
a non-linear crystal element into a coherent pare of two photons \cite{breitenbach}. An
interesting phenomenon has been observed in  the presence of squeezed states: the
fluctuations of the electric field are locally lower than the vacuum
fluctuations\footnote{This surprising conclusion is absolutely certain, as the squeezed
states even allowed for spectroscopic measurements \cite{kimble} that exhibit spectral
lines narrower than the natural line-width.}, the so-called shot-noise level. The amount
of this reduction, the so-called degree of squeezing, has been the subject of intensive
experimental studies. Recently this reduction has been pushed up to $-6.2dB$
\cite{schneider}.

As far as we know  quantum field theoreticians do not know that their inequalities may
influence real experiments nor are quantum opticians  aware of the existence of such
inequalities.

An immediate question arises, namely whether there is an analogue of a quantum
inequality for the reduction of the electric field fluctuations, and an even
deeper one of whether the reduction of fluctuations has anything in common with
negative energy densities. We shall give an affirmative answer to the first
question and an argument with regard to the second\footnote{A connection
between negative energy densities and the squeezed states has already been
mentioned in a popular article by L.Ford and T.Raman \cite{fordraman}.}.

Our paper is organized as follows: the second chapter contains a derivation of the
inequality for the reduction of the electric-field fluctuation. The inequality is
presented in the most general form with an arbitrary time-probe function. The reduction
is expressed, as typically for quantum inequalities, in units of energy. The third
chapter contains three adaptations of the inequality in the context of quantum-optical
experiments. Firstly, the observables are restricted in frequency, reflecting the
physical situation where all detectors are characterized by a frequency-dependent
sensitivity function, $\mu(\w)$. With this step the vacuum fluctuations become  finite.
Secondly, the maximal possible reduction is expressed in the $dB$ scale (minimal
fluctuations versus vacuum fluctuations) making it compatible with the language in which
experimental results are typically presented. Thirdly, two time-probe functions are
discussed, each dependent on the time-parameter $t_0$ expressing the length of the
interval in which the fluctuations are registered. Those three adaptations allow for a
prediction on the maximal degree of squeezing which is given in chapter four. In order
to establish this prediction we give supplementary arguments regarding the
interpretation of the local oscillator phase $\theta$ in the balanced homodyne detection
of the squeezed light. The first of the appendices contains a rather technical, but
elementary typical derivation of quantum inequalities. The second appendix presents an
actual calculation of the claimed maximal reductions of field fluctuations.

\section{Inequality for the reduction of electric field fluctuations}
As we shall see in the following investigations the method typically utilized in quantum
inequalities can easily be applied to certain observables which measure "the amount of"
electric field fluctuations in quantum optics. We shall prove  an inequality for
fluctuations of the electric part of the electromagnetic field similar to the one
obtained by Fewster and Teo in \cite{fewster} and recently by Pfenning \cite{pfenning}
(although those authors consider a much more complicated case of background
gravitational field). More precisely we will be interested in the expectation value of
the square of the electric field:
\begin{equation*}
  \langle \v E^2(t,\v x)\rangle.
\end{equation*}

The square of the electric-field operator is not a well-defined observable\footnote{It
does give infinite expectation values even in the vacuum state. It does not help if we
smooth it out by means of an integration with smooth functions of $t$ and $\v x$. In
short - $E^2(x)$ is not an operator valued distribution.}. In order to avoid mathematical
nonsense we must therefore define the observables of interest with care.

\subsection{Observable of interest} A precise definition of the square of the electric field is
provided by the standard procedure of point splitting. In the present context, where the
ground state is simply the vacuum state (denoted by $\W$), the point splitting results in
normal ordering.  The point splitting procedure gives a physical meaning to the normal
ordering (instead of regarding it as a dumb rule of mechanical operator ordering) which
we shall use later. We therefore recall it briefly:
\begin{itemize}
\item the bi-local observable $\v E^2(x,y)=\v E(x)\v E(y)$, is a well defined observable
(operator valued bi-distribution),
\item the bi-local difference
\begin{equation*}
  \D \v E^2(x,y)=\v E^2(x,y)- \langle \v E^2(x,y)\rangle_\W \mathcal{I}
  \end{equation*}
where $\mathcal{I}$ is the identity operator and $\langle. \rangle_{\W}$ denotes an
expectation value in the vacuum state, is also an operator-valued bi-distribution, and
\item the limit
\begin{equation*}
  \lim_{x\rightarrow y} \D \v E^2(x,y)\equiv\D \v E^2(x)=:\v E^2(x):
\end{equation*}
exists as an operator valued distribution.
\end{itemize}

Although $:\v E^2(x):$ is simply the normal ordering of the square of $\v E(t,\v x)$ we
see that its expectation value in a certain state $|S\rangle$ of the electromagnetic
field gives the difference of the expectation values of the squares of the field between
the state $|S\rangle$ and the vacuum state although those expectation values alone are
infinite and thus have no meaning.

\subsection{Inequality for the reduction of fluctuations}
In the following we will investigate the time-weighted electric field squares at a certain
point in space (say $\v x$):
\begin{equation}\label{proper_inequ}
\D=\int_{-\infty}^\infty dt \ f(t)  \ \D E^2(t,\v x).
\end{equation}
Here $f(t)$ denotes a positive, real-valued function which is normalized in the
probabilistic sense:
\begin{equation*}
  \int_{-\infty}^\infty f(t)\ dt=1.
\end{equation*}

We emphasize that the following considerations will be state-independent. We shall find
an inequality which is necessarily  fulfilled by the quantized radiation in any
physically allowed state (be it a coherent, multi-photon, squeezed, thermal or any other
state).

We quantize the radiation field in the Coulomb gauge:
\begin{align*}
A_0&=0 & \d_i A^i&=0
\end{align*}
finding the standard expression for the potentials:
\begin{equation*}
  A_i(x)=\frac{1}{\sqrt{2\pi}^3} \int \frac{d^3k}{\sqrt{2\w_k}}\ \sum_{\a=1,2}\v e^\a_i(k)
  \left\{\he a_\a (k) e^{ikx}+a_\a (k) e^{-ikx}\right\},
\end{equation*}
where $\w_k=|\v k|$ and $(\v e^\a)$ denote the two real valued polarization vectors
(labeled by $\a$) which are normalized and orthogonal to $\v k$. The electric field
operator is found by means of
\begin{equation*}
  E_i=-\d_tA_i.
\end{equation*}
The point-splitting (normal-ordered) operator of the squared electric field is
\begin{multline*}
  \D E^2(t,\v x)=\frac{1}{(2\pi)^3} \int \frac{d^3k \ d^3p}{2\sqrt{\w_k\w_p}}\ \w_k\w_p
  \sum_{\a,\be=1,2} \v e^\a(k) \v e^\be(p)\\
  \left\{\he a_\a (k) a_\be(p)e^{i(k-p)x}-a_\a (k) a_\be (p) e^{-i(k+p)x}
  +h.c.\right\},
\end{multline*}
where $\v e^\a(k) \v e^\be(p)$ denote the 3D scalar product of the two polarization
vectors\footnote{The 4D and 3D products are only distinguished by means of their
arguments i.e. bold face vectors are three-dimensional whereas the normal-typed are 4D.
For 4D products we use the convention $x y= x_0 y_0 - \v x \v y$.}. In order to obtain
the observable $\Delta$ we integrate the above expression with $f(t)$, while using the
following convention for the Fourier transform:
\begin{equation*}
  \frac{1}{2\pi}\int dt \ e^{-i\w t}f(t)=\hat f(\w).
\end{equation*}
We obtain
\begin{multline}\label{Delta_E}
  \D=\frac{1}{2(2\pi)^2} \int {d^3k \ d^3p}\ \sqrt{\w_k\w_p}\ \sum_{\a,\be=1,2}\v e^\a(k) \v e^\be(p)\\
  \left\{\he a_\a (k) a_\be(p)e^{i(-\v k+\v p)\v x}\hat f(\w_p-\w_k) -
  a_\a (k) a_\be (p) e^{i(\v k+\v p)\v x}\hat f(\w_k+\w_p) +h.c.\right\}.
\end{multline}

In order to find a state-independent inequality for the expectation value of $\Delta$ we proceed
analogously to the scalar-field case (inequality \eqref{inequality}) presented in the appendix
\ref{estimate}. Here, in the case of electromagnetic (vector) field a slight modification is
necessary in order to take care of the polarization vectors $\v e^\a(k)$. We define vector
operators $\v B$:

\begin{equation}
  B^i(\w)=\int d^3p \ \sum_{\a=1,2} \v e^i_\a(p) \left\{\ba{g(\w-\w_p)}\chi(p) a_\a(p)
  -
  \ba{g(\w+\w_p})\ba{\chi(p)}\he a_\a(p)\right\},
\end{equation}
and we investigate the positive operator:
\begin{equation}\label{babar}
  \int_0^\infty  d\w \ \he B(\w)^i B(\w)^j \ {\de_{ij}}.
\end{equation}
In the calculation, completely analogous to that  presented in the appendix
\ref{estimate} for a scalar-field case, the commutation relations:
\begin{equation*}
  [a_\a(p),\he a_\be(k)]=\de(\v p-\v k )\de_{\a\be},
\end{equation*}
lead to a factor
\begin{multline*}
  \int_0^\infty d\w \int_{\mathbb{R}^3} d^3p\
  \left|\widehat{\sqrt{f}}(\w+\w_p)\right|^2 |\chi(p)|^2 \ \v e^i_\a(p) \v e^j_\be(k)\
  \de_{ij}\de^{\a\be}=\\=
2\* \int_0^\infty d\w \int_{\mathbb{R}^3} d^3p\
  \left|\widehat{\sqrt{f}}(\w+\w_p)\right|^2 |\chi(p)|^2,
\end{multline*}
which is the essence of the inequality we were searching for. Here, the factor $2$ is the
only remainder of the polarization vectors. We obtain an identity (an analogue of
\eqref{equality}):
\begin{multline}\int_0^\infty  d\w \ \he B(\w)^i B(\w)^j \ {\de_{ij}}=\frac{1}{2}\int d^3p
\ d^3k\  \sum_{\a,\be=1,2} \v e^\a(p) \v e^\be(k)\* \\ \left\{  \he a_\a(p) a_\be(k)\
\ba{\chi(p)}\chi(k) \hat f(\w_k-\w_p)- a_\a(p) a_\be (k) \ {\chi(p}) \chi(k) \hat f(\w_k+\w_p)+ h.c.\right\}+\\
  +2\int_0^\infty d\w \int_{\mathbb{R}^3} d^3p\
  \left|\widehat{\sqrt{f}}(\w+\w_p)\right|^2 |\chi(p)|^2.
\end{multline}
By means of  comparison with \eqref{Delta_E} we find:
\begin{equation}\label{ident}
  \chi(p)=\frac{e^{i\v p \v x} \sqrt{\w_p}}{\sqrt{2\pi}^{\ 2}}.
\end{equation}
The positivity of the operator \eqref{babar} leads, after taking an expectation value in
some (arbitrary) state of the field $|S\rangle$, to the following inequality:

\begin{equation}\label{the_result}
\int_{-\infty}^\infty dt \ f(t) \langle \D E^2(t,\v x)\rangle_S\geqslant
  -\frac{2}{(2\pi)^2}\int_0^\infty d\w \int_{\mathbb{R}^3} d^3p\
  \left|\widehat{\sqrt{f}}(\w+\w_p)\right|^2 \w_p.
\end{equation}

\section{Consequences of the fluctuation-inequality}
The main result of this paper, the inequality \eqref{the_result}, tells us that the
vacuum fluctuations can be reduced only in a limited way. The function $f(t)$ on the one
hand specifies the time interval in which fluctuations are recorded and on the other hand
limits the amount of the reduction of the vacuum fluctuations one  will be allowed to
record. In the following we will investigate the practical consequences of the
established inequality.

Although the inequality gives an absolute value, in the units of energy, of the maximal
reduction of the vacuum fluctuations, it is of little importance since it gives a finite
number, the reduction, which is cut from an infinite reservoir (because the vacuum
fluctuations are infinite). What seems to have a far deeper practical consequence is
another quantity, namely, the reduction of the fluctuations restricted to certain
frequencies. All available detectors of the radiation, be it single atoms, photodiodes
or any other device, are always sensitive only to certain frequencies of the radiation.
Let $\mu(\w_P)$ denote a smooth sensitivity function of rapid decay (in short $\mu_p$).
If we exchange the measure $d^3p\rightarrow \mu_p \ d^3p$ in order to take into account
this physical restriction, then even  the vacuum fluctuations become finite:

\begin{multline}\label{vacuum}
\langle E^2\rangle_\Omega:=\int_{-\infty}^\infty dt \ f(t) \langle E^2(t,\v x)\rangle_\Omega=\\
= \frac{1}{2(2\pi)^3}    \int \mu_k \ d^3k \ \mu_p \ d^3p\  \sqrt{\w_k\w_p}
\sum_{\a,\be=1,2} \v e^\a(k) \v e^\be(p) \de_{\a\be} \
 \de(\v p- \v k) =\frac{1}{(2\pi)^3}\int (\mu_p)^2 \ d^3p \ \w_p.
\end{multline}
The inequality \eqref{the_result} is also modified by means of the
additional factor $(\mu_p)^2$ appearing on its right-hand side.

Of prime interest is the logarithmic reduction of the fluctuations defined as the
base-10 logarithm of the field fluctuations in the considered state $|S\rangle$ divided
by the vacuum-fluctuations (the reduction in [dB]):
\begin{equation*}
  R:=10 \cdot \log_{10}\left(\frac{\langle E^2\rangle_S}{\langle E^2\rangle_\Omega}\right)=
  10 \cdot \log_{10}\left(\frac{\langle \D \rangle_S+\langle E^2\rangle_\Omega}{\langle
  E^2\rangle_\Omega}\right),
\end{equation*}
where in the presence  of the sensitivity function $\mu_p$
\begin{equation}\label{def_delta}
  \langle \D\rangle_S=\langle E^2\rangle_S-\langle E^2\rangle_\Omega
\end{equation}
makes sense with both quantities on the right hand side finite. We shall compute this
quantity for  certain model time-probe functions $f(t)$.

\subsection{Time probe functions}
The probe function $f(t)$ was until now arbitrary. Nevertheless, it will be instructive
to consider some examples which shed  additional light on the meaning of the proven
inequality.

Let us consider the following function:
\begin{equation*}
  f(t)=\frac{2}{\pi}\frac{t_0^3}{(t^2+t_0^2)^2}.
\end{equation*}
The function is even, normalized as the probability density and has its maximum at the
zero $f(0)={2}/(\pi t_0)$. Its half-width is connected linearly with $t_0$ c.a.
$t_{h-w}\approx 0.6 \cdot t_0$. The second power of the Fourier transform of the square
root of $f(t)$ is easily found to be
\begin{equation*}
  \left(\widehat{\sqrt{f(t)}}\right)^2(\w)=\frac{t_0}{2\pi}e^{-2|\w|t_0}.
\end{equation*}
We are now in a position to apply the proven inequality. According to \eqref{the_result}
we find
\begin{equation}\label{delta_max}
  \langle \D\rangle\geqslant -\frac{1}{(2\pi)^3}\int(\mu_p)^2\  d^3p \ \w_p \ e^{-2pt_0}=
  :\langle \D\rangle_{max}.
\end{equation}
The above result, when compared with the vacuum fluctuations $\langle
E^2\rangle_\Omega$, shows the merit of the fluctuation-inequality \eqref{the_result},
namely although the minimal fluctuations at an instant of time\footnote{In the limit
$t_0\rightarrow0$ the sensitivity function becomes the delta distribution $\de(t)$.} are
simply zero (compare with \eqref{def_delta}),
\begin{equation*}
  0\geqslant\langle \D\rangle_{max}\geqslant-\langle E^2\rangle_\Omega,
\end{equation*}
such a reduction (according to \eqref{delta_max}) cannot last for a finite period of time. The
longer the probe time becomes (bigger $t_0$) the bigger the square of the field $\langle
E^2\rangle_S$ have to be expected for all states of the electromagnetic field.

There is a certain objection which can be made against the probe function we  used in
the above considerations, namely, an experimentator would expect  to be able to produce
huge reductions of fluctuations for only relatively short periods of time only after
which he would expect a period of greatly increased $\langle E^2\rangle_S$. Thus, he
would not be satisfied with the fact that the probe function decays only with the
inverse fourth power of the time and would rather seek a more rapidly decaying function
so that the increased fluctuations do not influence the reduced ones.

Indeed, we regard the  point made above as a fair challenge and thus consider another
probe function, namely the  gaussian function
\begin{equation}\label{gauss}
  g(t)=\frac{1}{t_0\sqrt{2\pi}} \ e^{-\frac{t^2}{2t_0^2}},
\end{equation}
which certainly possesses much better time-decay properties. Noting that
\begin{equation*}
 \left(\widehat{\sqrt{g(t)}}\right)^2(\w)=\frac{t_0}{\pi \sqrt{2\pi}} \ e^{-2\w^2t_0^2}
\end{equation*}
we find
\begin{equation}\label{maximum}
   -\frac{2}{(2\pi)^2}\frac{2 t_0}{2 \pi \sqrt{2\pi}}
  \int_0^\infty d\w \int(\mu_p)^2\  d^3p \ \w_p \ e^{-2(|p|+\w)^2t_0^2}=
  :\langle \D\rangle_{max}.
\end{equation}
Now, due to
\begin{equation*}
  \frac{4t_0}{\sqrt{2\pi}}\int_0^\infty d\w \ e^{-2(|p|+\w)^2t_0^2}\leqslant 1
\end{equation*}
which holds for all $|p|>0$ (and the equality is approximate in the limit $t_0\rightarrow 0$)
the same remarks as in the case of previously investigated time-probe function $f(t)$ namely
\begin{equation*}
  0\geqslant\langle \D\rangle_{max}\geqslant-\langle E^2\rangle_\Omega,
\end{equation*}
can be expressed, so that the total suppression of $\langle E^2\rangle_S$ is possible
only in the limit\footnote{We should warn however that we have nothing to say about the
limit $t_0\rightarrow 0$ for arbitrary time-probe functions. Indeed this would be
equivalent  to giving a meaning to the square root of the delta distribution which
however we fear of.} $t_0\rightarrow 0$.

In the following investigations we shall try to relate the inequality to the measured
degree of squeezing in quantum-optical experiments.

\section{Squeezed states on BHD and inequalities for the degree of squeezing}

Suppose a beam  of squeezed light is shed upon a balanced homodyne
detector\cite{homodyne} (BHD). Putting away for a moment the theoretical description of
the squeezed states in terms of the creation/annihilation operators we present an
experimentally motivated argument which allows for an application of the inequality.

Firstly it appears possible to phase match the squeezed light to the local oscillator
field and to manipulate the relative phase $\theta$ in a controlled way\cite{zhang}. As
the phase $\theta$ is typically manipulated by means of moving back and forth a
piezo-mounted mirror it appears plausible that $\theta$ should be regarded as a
time-delay\footnote{If the local oscillator were e.g. pulsed \cite{mc_alister,hansen}
than $\theta$ would move (delay) the pulse back and forth in time further strengthening
our interpretation.}. As a byproduct we note that if a certain aparatus (e.g. BHD) were
to measure the field operator restricted to a sharp frequency $\w$ then the
transformation $\theta\rightarrow \theta + \pi/2$ would exchange the meaning of the
so-called field quadratures:
\begin{equation*}
(a(\w)+\he a(\w))/2 \equiv E_1(\w)\rightarrow E_2(\w)\equiv(a(\w)-\he a(\w))/2.
\end{equation*}
From now on we shall therefore regard $\theta$ as a time-delay.

\begin{figure}[htb]
\center
  \includegraphics{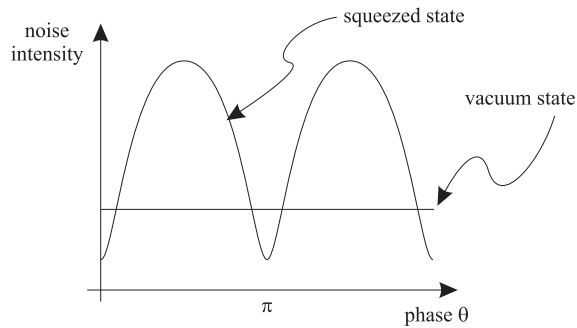}
  \caption{Typical dependence of field fluctuations on the phase of a local oscillator. Intervals
  where the fluctuations dive under vacuum fluctuations correspond to squeezing.}
  \label{plot}
\end{figure}

Secondly the squeezed states exhibit a typical $\theta$-dependence of the field's
fluctuations (see figure \ref{plot} or e.g. \cite{kimble_sq_gen,zhang}). In accordance
with the above interpretation, $\theta=0$ corresponds to a measurement of the field (and
its variation - its square) at a certain instant of time $t$, whereas $\theta=\pi/2$
corresponds to a measurement at a later time $t+T/4$, where $T$ is the period of the
local oscillator. If the variations of the measurement at $\theta=0$ are small and those
at $\theta=\pi/2$ are huge, as on the figure, we are inclined to believe the periods of
huge fluctuations come directly ($T/4\approx 10^{-15} s$) after those of small
fluctuations.

Furthermore all possible evidence (e.g. \cite{breitenbach}) supports the fact, that the
squeezed state consists of a superposition of pairs of photons. Thus, whatever the
Fock-state representation of the squeezed state $|S\rangle$ is, it only contains vectors
with even photon number. Consequently,
\begin{equation*}
  \langle \v E(t,\v x)\rangle_S=0, \quad \forall t, \v x
\end{equation*}
as the field operator $\v E(t,\v x)$ contains only one creation and one annihilation
operator so that the above relation follows from the commutation relations. We therefore
see that field's fluctuations (variation of the field measurement) are directly related
to the expectation value of the square of the field.

Given experimental precision in the control of the phase we can safely assume
that the most intense squeezing lasts for $T/100$. Experimentally reported
continuous squeezing for much longer periods \cite{zhang} although seemingly
contradictory the above statement uses the fact the BHD reacts only in correct
moments, selected by the local oscillator and therefore is sensitive only to
 certain sub-intervals within each period preferably to those of maximum squeezing.

The above interpretation allows finally for an application of the inequality
\eqref{the_result}. Choosing the characteristic time interval $t_0=T/100$ and
frequency range $\mu_p$ (see footnote in appendix \ref{example}) of interest we
can estimate the maximum possible reduction of the vacuum fluctuations. Here
the concern about the fast-decrease property of the probe function becomes
visible as the intervals of anti-squeezing come immediately after the intervals
of squeezing. Nonetheless, the application of the inequality reveals (see
appendix \ref{example}) in the case of gaussian probe-function the best
possible squeezing lasting for $T/100$:
\begin{equation*}
  max\ squeezing= R(0.01) = -14.96 \ dB.
\end{equation*}

The above result might have a direct impact on  experiments in the near future limiting
further progress in the creation of strongly-squeezed states. It is however important to
stress that the limit binds the maximal reduction to the length of the time interval of
its appearance. There is no reason not to be able to squeeze up to $-25 \ dB$ if only the
phase $\theta$ could be controlled with $0.1\%$ precision.

In short: the deduction of the above nontrivial limitation  contains three ingredients:
\begin{enumerate}
\item the fluctuation inequality \eqref{the_result},
\item the assumption that the squeezed state contains pairs of photons and thus $\langle E\rangle_S=0$,
\item the interpretation that $\theta$ has a meaning of time delay so that periods of squeezed
field come directly after periods of anti-squeezed field.
\end{enumerate}

The first ingredient should, in the light of presented calculations, be regarded as a
mathematically certainty; its validity is as certain as the description of light in terms
of quantized field. The second ingredient appears physically certain as it is supported
by overwhelming physical evidence. The third ingredient is supported by the above
arguments. In our opinion it gives a correct interpretation of the so-called field
quadratures.

Let us again stress that the third ingredient specifies very short times for which the
field is probed (i.e. $t_0$). Even if our interpretation of $\theta$ appears odd it is
important to understand that there is little freedom left; namely in  light of the
fluctuation inequality it is impossible to claim that $\langle E^2\rangle_S$ was kept at
say $-6\ dB$ even for one period $T$ (the maximum allowed degree of squeezing for
$t_0=T$ is $R(1)=-.00027 \ dB$) so that it would be contradictory to claim that the
experimentally reported squeezing lasted longer than a fraction of $T$ .

\section{Remarks and outlook}\label{outlook}

A number of important issues have been left to the present section:
\begin{itemize}
\item It should be firmly stressed that the presented inequality has hardly anything in common with the
standard inequality
\begin{equation*}
   \langle \D E_1^2 \rangle\langle \D E_2^2\rangle\geq \hbar,
\end{equation*}
typically present in the discussion of squeezed states, where  $E_1$ denotes one field
quadrature, say squeezed, while $E_2$ the other, anti-squeezed. Indeed, if a definite
time interval is considered (say, $t_0$) then the above inequality does not prohibit an
arbitrarily small fluctuations $\langle \D E_1^2\rangle$ in that interval with the
consequence that the fluctuations $\langle \D E_2^2\rangle$ in the meantime explode. On
the other hand, the presented inequality prohibits exactly that. A rough statement would
be that in the light of the presented inequality $\langle \D E_1^2\rangle$ alone is
restricted from below.

\item The presented inequality possesses a clear spatio-temporal interpretation. If there were any
hopes to use squeezed light in order to inhibit atomic phase decays and make some
relaxation times arbitrarily long \cite{gardiner} they must be seriously reduced in
light of the presented inequality.

\item The most important question, whether squeezed states exhibit periods of negative energy
densities remains to be answered. As the energy-density of the electro-magnetic field
contains a square of the magnetic field:
\begin{equation*}
  \p(x)=\frac{1}{2}(E^2(x)+B^2(x)),
\end{equation*}
the question arises whether magnetic field fluctuations are also suppressed and whether
the periods of their suppression coincide with periods of squeezing. As magnetic field
fluctuations are not typically recorded in experiments we endeavor to infer the answer
theoretically.
    Assume the squeezed field is linearly polarized as is the case in some experiments
\cite{kimble_sq_gen}. Furthermore assume that the wavepackets of photons constituting the
pair \eqref{state} are compactly localized along certain  momentum-vector $\v k_x$ and
propagate in one direction only (see figure \ref{supp}). The only difference between
$E^2(x)$ and $B^2(x)$ is that instead of $\w_k\w_p\ \v e(\v k) \v e(\v p)$ there appears
a factor $[\v k \times \v e(\v k)] [\v p \times \v e(\v p)]$.

\begin{figure}[htb]
\center
  \includegraphics{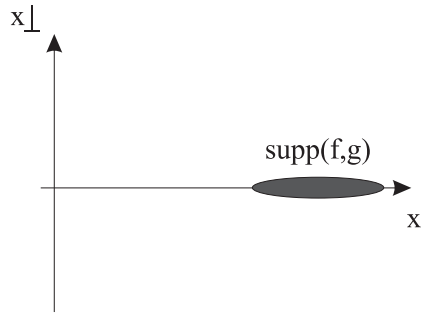}
  \caption{Support of photons wavepackets in momentum space.}
  \label{supp}
\end{figure}

If the expectation value of the normal ordered $B^2(x)$ is calculated then the above
factor is integrated with the wavepackets. It is therefore allowed  to restrict the $k$
and $p$ integration to the support of the wavepackets from the very beginning. Then each
of the vectors may be decomposed orthogonally into $\v k= \v k_x + \v k_\bot$ and $\v p=
\v p_x + \v p_\bot$ so that $\v k_x$ is parallel to $\v p_x$. The assumption on the
support of the wavepackets tells us that $|\v k_x|$ is much greater than $|\v k_\bot|$. A
calculation shows that
\begin{equation*}
[\v k \times \v e(\v k)] [\v p \times \v e(\v p)]=\w_k\w_p\ [\v e(\v k) \v e(\v p)+ O(\w^{-2})],
\end{equation*}
so that the expectation value of $E^2(x)$ is approximately equal to that of $B^2(x)$.

As we have argued the periods of squeezing should coincide with the periods of negative
energy density. However, squeezing is typically observed with restriction to a sharp
frequency. Only recently \cite{breitenbach} the so-called broadband detection of squeezed
light was performed. Although the squeezing at many different frequencies was indeed
observed it is not clear to us whether there existed periods where squeezing occurred in
all frequencies simultaneously which would be necessary in order to infer that the
negative energy densities were indeed created experimentally. If it were so, the squeezed
states would provide the first known experimental example of locally negative energy
densities.
\end{itemize}

\section{Acknowledgments}
It is a pleasure to thank K. Fredenhagen for his encouraging remarks as well as
Graduiertenkolleg "Zuk\"unftige Entwicklungen in der Teilchenphysik" for financial
support.

\appendix
\section{Inequality for scalar field}\label{estimate}
In the present appendix we outline a standard method associated with the
derivation of quantum inequalities. We consider the scalar-field case in which
the derivation already contains all essential features.

Let us give an outline of this derivation. We shall consider the operator
$A(\chi,f)$ which corresponds to the time weighted normal ordered square of the
electric field (i.e. to the observable $\D$ see equation \eqref{proper_inequ}
and \eqref{Delta_E}). It will be shown that
\begin{equation*}
    \langle A(\chi,f) \rangle \geqslant - \text{positive functional of $\chi$ and $f$}
\end{equation*}
where the functional on the RHS is finite if the time-probe function is fast
decreasing. The proof consists of finding an appropriate operator $B(\w)$,
where $\w$ is real number (the operator also depends on $\chi$ and $f$) such
that
\begin{equation*}
    \int_0^\infty \he B (\w) B(\w) d\w=A(\chi,f)+\text{residue term},
\end{equation*}
where the residue term is precisely the functional described above. Although
$A(\chi,f)$ will prove to be just the normal ordering of $\int_0^\infty \he
B(\w) B(\w) d\w$ it should perhaps be stressed that the derivation has nothing
to do with normal ordering of the square of the field operators $ E(x) E(y)$ in
which case the residue term would prove to be infinite and the inequality would
be trivial i.e. formally $\langle E^2\rangle\geqslant 0$. We now proceed with
the scalar field derivation.

The operator $A(\chi,f)$ corresponding to \eqref{Delta_E} is defined as:
\begin{multline}\label{operator_A}
  A(\chi,f)=\frac{1}{2}\int d^3p \ d^3k\  \left\{  \he a(p) a(k) \ba{\chi(p})
  \chi(k) f(\w_k-\w_p)\right.
  +\Bigl. a(p) a(k) {\chi(p}) \chi(k) f(\w_k+\w_p)\Bigr\}+ h.c.
\end{multline}
and will be shown to fulfill the inequality:
\begin{equation}\label{inequality}
  \langle A(\chi,f)\rangle_S\geqslant- \int_0^\infty d\w \int_{\mathbb{R}^3} d^3p\
  \left|\widehat{\sqrt{f}}(\w+\w_p)\right|^2 |\chi(p)|^2.
\end{equation}
 in any state $S$ of the quantum field. We prove this by finding an operator $B(\w)$ such that
\begin{equation}\label{int_bb}
  \int_0^\infty  d\w \ \he B(\w) B(\w)= A(\chi,f)+\int_0^\infty d\w \int_{\mathbb{R}^3} d^3p\
  \left|\widehat{\sqrt{f}}(\w+\w_p)\right|^2 |\chi(p)|^2.
\end{equation}
Obviously the inequality follows since $\int d\w \ \he B(\w) B(\w)$ is a positive
operator.

Define
\begin{displaymath}
  B(\w)=\int d^3p \ \left\{\ba{g(\w-\w_p)}\chi(p) a(p) + \ba{g(\w+\w_p})\ba{\chi(p)}\he
  a(p)\right\},
\end{displaymath}
where $g(x)$ is a positive square root of $f(x)$ i.e. its Fourier transform fulfills
\begin{align*}
\ba{g(p)}&=g(-p)\\
\int_{-\infty}^\infty d\w \ g(p-\w) g(\w)&=f(p).
\end{align*}
Then \eqref{int_bb} is investigated:
\begin{multline}
\int_0^\infty d\w \ \he B(\w) B(\w)= \int_0^\infty d\w \ d^3p\  d^3k\
\left\{ g(\w-\w_p) \ba{g(\w-\w_k})\ba{\chi(p)}\chi(k) \he a(p)
a(k)+\right.\\ + \left. g(\w+\w_p) \ba{g(\w+\w_k})\chi(p)\ba{\chi(k)}
a(p) \he a(k)+\right.  \left. g(\w-\w_p)
\ba{g(\w+\w_k})\ba{\chi(p})\ba{\chi(k)} \he a(p) \he a(k)+\right.\\+
\left. g(\w+\w_p) \ba{g(\w-\w_k}){\chi(p)}{\chi(k)} a(p) a(k) \right\}
\end{multline}

Let us refer to the components above as $I,II,III,IV$. In the second
($II$) term we use the commutation relations and obtain
\begin{multline}
  II=\int_0^\infty d\w \ d^3p\  d^3k\  g(\w+\w_p)
\ba{g(\w+\w_k})\chi(p)\ba{\chi(k)}  \he a(k) a(p)+\int_0^\infty d\w \int
d^3p \left|g(\w+\w_p)\right|^2 |\chi(p)|^2.
\end{multline}
Now the first part of the above operator (denoted by $II'$), together with the first term ($I$)
will be investigated. In $II'$ we exchange the variables $p\leftrightarrow k$, make a further
change $\w \rightarrow -\w$ and use the relation $\ba{g(-\w+\w_p)}=g(\w-\w_p)$. As a result we
obtain
\begin{displaymath}
  I+II'=\int_{\-\infty}^\infty d\w \ d^3p \ d^3k \ g(\w-\w_p) g(\w_k-\w) \ba{\chi(p)}\chi(k)
  \he a(p) a(k)
\end{displaymath}
The integration in $\w$ may be recognized as a convolution so that
\begin{displaymath}
  I+II'=\int d^3p \ d^3k \ f(\w_k-\w_p) \ba{\chi(p)}\chi(k)
  \he a(p) a(k).
\end{displaymath}
After a similar treatment the terms $III$ and $IV$ are transformed\footnote{In the
calculation one first splits the expressions into two equal parts and than performs
$\w\rightarrow-\w$ in one of the parts, which after subsequent exchange
$p\leftrightarrow k$ allows to stretch the $\w$ integration to the whole $\mathbb{R}$.}
into
\begin{align*}
  III&=\frac{1}{2}\int d^3p \ d^3k \ \ba{f(\w_k+\w_p}) \ba{\chi(p})\ba{\chi(k})
  \he a(p) \he a(k)\\
  IV&=\frac{1}{2}\int d^3p \ d^3k \ f(\w_k+\w_p) {\chi(p)}\chi(k)
  a(p) a(k).
\end{align*}
Altogether we have
\begin{multline}\label{equality}
\int_0^\infty \he B(\w) B(w)= \frac{1}{2}\int d^3p \ d^3k \left\{f(\w_k-\w_p)\ba{\chi(p})
\chi(k)\ \he  a(p) a(k)  + \right.\\
+\Bigl. f(\w_k+\w_p) {\chi(p)}\chi(k)
 \  a(p) a(k)+h.c.\Bigr\} + \int_0^\infty d\w \int d^3p
\left|g(\w+\w_p)\right|^2 |\chi(p)|^2.
\end{multline}
The desired inequality \eqref{inequality} now follows easily by taking
the expectation value. It should perhaps be firmly stressed that this
inequality is universal in that it must be respected by all states
$|S\rangle$ of the quantum field.

As a corollary we note that the sign in front of the $a(p)a(k)$ term may be changed:
\begin{multline}
  \tilde{A}(\chi,f)=\int d^3p \ d^3k\  \left\{
  \he a(p) a(k) \ba{\chi(p}) \chi(k)
  f(\w_k-\w_p)- a(p) a(k) {\chi(p}) \chi(k) f(\w_k+\w_p)\right\}+ h.c.
\end{multline}
and still the same inequality would be obtained if only the operator
\begin{displaymath}
  \tilde{B}(\w)=\int d^3p \ \left\{\ba{g(\w-\w_p)}\chi(p) a(p) - \ba{g(\w+\w_p})\ba{\chi(p)}\he
  a(p)\right\}
\end{displaymath}
is utilized instead of $B(\w)$.

\section{Restriction on the degree of squeezing\label{example}}

Here we shall briefly indicate what sort of limitations on the degree of
squeezing are to be expected. We take the gaussian probe function \eqref{gauss}
and assume a sharp frequency cut so that the function  $\mu(\w_p)$ is sharply
centered around certain fixed frequency\footnote{Typically the field is
measured on BHD with local oscillator LO \cite{breitenbach} (mainly because no
device is sensitive enough to measure the vacuum fluctuations directly). The
frequency of the LO plus the frequency at which the output of BHD is analyzed
give the center $\w_0$ of the sensitivity function $\mu_p$. The bandwidth of
the BHD output measurement defines the shape of $\mu_p$. In the following the
bandwidth is assumed to be negligibly small with respect to $\w_0$. This
assumption allows us to derive explicit result and but is by no means
necessary.} $\w_0$. Then the maximum reduction of fluctuations is given by
\eqref{maximum}. Together with the vacuum fluctuations \eqref{vacuum} the
following maximum degree of squeezing results:
\begin{equation*}
  R(\tau)=10\cdot \log\left[1-\frac{4}{\sqrt{2\pi}}\int_0^\infty ds \ e^{-2(s+\tau)}\right],
\end{equation*}
where $\tau=\w_0 t_0$. Introducing the error function $\text{erf}(x)=\frac{2}{\sqrt{\pi}}\
\int_0^x e^{-t^2} \ dt$ we find
\begin{equation*}
  R(\tau)=10 \cdot \log[\text{erf}(2\sqrt{2} \tau)],
\end{equation*}
which is the maximum reduction of fluctuations allowed by the inequality
\eqref{the_result}. If, as in typical experiments with squeezed light, the period of
maximum squeezing were to last for $1\%$ of the period $T$ (i.e. $\tau=0.01$), the
squeezing cannot exceed
\begin{equation*}
  R(0.01)=-14.96 \ dB.
\end{equation*}
If on the other hand someone would insist the squeezing lasted longer - say the whole
period $T$ or even more, it would be bounded by
\begin{equation*}
  R(1)=-.00027 \ dB.
\end{equation*}


\end{document}